\documentclass[preprint,onecolumn,nofootinbib,superscriptaddress]{revtex4-1}

\usepackage{lineno}
\usepackage{braket}
\usepackage{mathtools}
\usepackage{ulem}

\newcommand{\be}{\begin{equation}}
\newcommand{\ee}{\end{equation}}

\usepackage{color}
\definecolor{purple}{rgb}{.36,.12,.60}
\definecolor{orange}{rgb}{.9,.3,.0}

\begin{document}

\title {
%v1 Non-Diffractive Polarization Properties of Optical Vortices \\ v2 Propagation-Invariant Polarization Features of Optical Vortex Beams
Non-Diffractive 3D Polarisation Features of Optical Vortex Beams}

\author{Andrei Afanasev}

\affiliation{Department of Physics,
The George Washington University, Washington, DC 20052, USA}
\author{Jack J. Kingsley-Smith}

\author{Francisco~J.~Rodr\'iguez-Fortu\~no}

\author{Anatoly~V.~Zayats}

\affiliation{Department of Physics and London Centre for Nanotechnology, King's College London, Strand, London WC2R 2LS, UK}

%\affiliation{And ... ?}

\begin{abstract}
Vector optical vortices exhibit complex polarisation patterns due to the interplay between spin and orbital angular momenta. Here we demonstrate, both analytically and with simulations, that certain polarisation features of optical vortex beams maintain constant transverse spatial dimensions independently of beam divergence due to diffraction. These polarisation features appear in the vicinity of the phase singularity and are associated with the presence of longitudinal electric fields. The predicted effect may prove important in metrology and high resolution imaging applications. 

%which reveal propagation-independent magnitude ratio with respect to their transverse counterparts. 
\end{abstract}
\date{\today
}
\maketitle

%%%%%%%%%%%%%%%%%%%%%%%%%%%%%%%%%%

\section{Introduction}

The interplay between spin and orbital angular momenta of light beams results in complex polarisation textures of light fields with optical properties important in imaging, metrology and quantum technologies \cite{shen2022topological}. For example, polarisation variations appear in the structure of two-dimensional photonic spin-skyrmions at length-scales much smaller than the wavelength of light because, in contrast to the field and intensity variations, the polarisation structure is not influenced by diffraction of electromagnetic waves \cite{du2019deep}. Such polarisation features often appear due to the spin-orbit interactions involving vector vortex beams and in the case of the evanescent fields may be topologically protected by the optical spin-Hall effect \cite{bliokh2015spin}. For three-dimensional (3D) free-space beams, such topological protection is not ensured and the polarisation features may vary significantly  upon beam propagation. On the other hand, observations of enhanced robustness of polarisation inhomogeneities in 3D structured light have been reported \cite{nape2022revealing}. 
Polarisation singularities of optical fields and their relation to phase singularities were discussed in Ref.\cite{berry2001polarization} and became an active field of research; see  Ref.\cite{senthilkumaran2020phase} for a review of the most recent developments.

In this paper, we show that the transverse size of certain polarisation features of optical vortex beams is preserved independently of the diffraction of the beam. The effect is governed by the phase singularity in the cross-section of the beam and arises due to the interplay of the longitudinal and transverse electromagnetic fields in the vector vortex.
The important role of the longitudinal fields was previously emphasized in the literature in the context of twisted-photon absorption by atoms \cite{quinteiro2017twisted,Afanasev_2020} and optical vortex dichroism \cite{forbes2019spin,de2020photoelectric}. The paper is organised as follows. In Sec.~2, we derive analytic expressions for the longitudinal-to-transverse field ratio near the beam phase singularity, and demonstrate its independence on the beam waist in a paraxial limit. Section 3 introduces a formalism for 3D optical field polarisation and shows, in an analytic model, that the transverse spatial profiles of the polarisation features are independent of beam divergence due to diffraction and the beam focusing conditions, and are dependent on the topological charge of the beam. Finally, in Sec.~4, we use full wave simulations beyond the paraxial limit to show the diffraction-independent polarisation features and confirm the analytic results.

%%%%%%%%%%%%%%%%%%%%%%%%%%%%%%%%%%

\section{Optical Vortex Fields Near the Phase Singularity}

%{\color{orange} (Here we consider different versions of optical vortex beams and their polarization.  Relations between longitudinal and transverse fields from a continuity equation. Independence of L/T field ratio on beam waist).}
We initially consider a paraxial monochromatic Laguerre-Gauss beam with a topological charge $l$ and zero radial index propagating in the $z$-direction (see the Appendix for a general case).  The electric field components in the transverse ($xy$)-plane, with the position vector $\mathbf r=(\rho \cos\phi,\rho \sin\phi, z)$ in cylindrical coordinates, are given by
\begin{equation}
\label{eq:Eperp}
\mathbf{E}_\perp(\mathbf{r}) = \text{\boldmath$\eta_\perp$} A_\perp \frac{w_0}{w(z)} \bigg(\frac{\rho}{w(z)}\bigg)^{|l|}  e^{-\frac{\rho^2}{w^2(z)}} e^{i(l\phi + kz + k \frac{\rho^2}{2R(z)} -(l+1)\zeta(z))}.
\end{equation}
Here, the vector $\text{\boldmath$\eta_\perp$} = \eta_x \hat{\mathbf{x}} + \eta_y \hat{\mathbf{y}}$ defines the polarisation of the beam in the ($xy$)-plane ($\eta_x$ and $\eta_y$ are complex dimensionless scalars normalised such that $|$\text{\boldmath$\eta_\perp$}$|=1$), $A_\perp$ is a normalisation constant, $w_0$ is the beam waist at $z=0$,
$R(z)$ is the beam curvature radius, $\zeta(z)$ is the Gouy phase factor, and $w(z)=w_0\sqrt{1+\big(\frac{z}{z_R}\big)^2}$, where $z_R=\frac{kw_0^2}{2}$ in the Rayleigh length \cite{Saleh2007}. 
The longitudinal component $E_z$ can be found from the Maxwell's equation $\mathbf \nabla~ \cdot~ \mathbf E~=~0$. Using the paraxial condition $\frac{\partial E_z}{\partial z}=ik E_z$ results in the relation,
\begin{equation}
\label{eq:paraxial}
    \mathbf \nabla_{\perp}\cdot \mathbf{E}_\perp=-ik E_z.
\end{equation}
For a conventional Gaussian beam (Eg.~(\ref{eq:Eperp})) with $l=0$, it follows that 
\begin{equation}
\label{eq:ratiogaussian-x}
    \frac{E_z}{E_x}=\frac{i\rho\cos\phi}{z_R+\frac{z^2}{z_R}},    
\end{equation}
for the transverse field that is linearly polarised along the $x$-axis and, 
\begin{equation}
\label{eq:ratiogaussian}
    R_{LT}\equiv\frac{|E_z|}{|E_\perp|}=\frac{\rho}{z_R+\frac{z^2}{z_R}},   
\end{equation}
for a circularly polarised transverse field, where we introduced a longitudinal-to-transverse field magnitude ratio $R_{LT}$. In this case, for a non-vortex beam $l=0$, $R_{LT}$ is inversely proportional to $z_R$ and hence to the square of the beam waist in the focal plane, and falls off as $z^{-2}$ at large propagation distances $z$.

However, for a vortex beam with $l\ne 0$, choosing appropriate expressions for \text{\boldmath$\eta_\perp$} in Eq. (\ref{eq:Eperp}) for left-hand or right-hand circular  ($\sigma=\pm1$), linear, radial and azimuthal polarisation (see the Appendix), and keeping only lowest-power terms in $\rho$ -- $i.e.$ assuming $\rho \ll w(z)$ -- we obtain the following longitudinal-to-transverse field ratios
\begin{eqnarray}\label{eq:Rltcases}
R_{LT}(\zeta)= 
\begin{cases}
\sqrt{2}|\zeta|,\ \ \text{circular}\  (\sigma\cdot l< 0)\ \text{polarisation}\\
0,\ \ \ \ \ \ \ \ \ \   \text{circular}\ (\sigma\cdot l>0) \ \text{polarisation}\\
|\zeta|,\ \ \ \ \ \ \text{linear polarisation}\\
2|\zeta|,\ \ \ \   \text{radial polarisation}\\
0,\ \ \ \ \ \ \ \ \ \text{azimuthal polarisation},
\end{cases} 
\end{eqnarray}
where we defined a quantity $\zeta \equiv l/(k \rho)$. Unlike the Gaussian beam results in Eqs.~(\ref{eq:ratiogaussian-x},\ref{eq:ratiogaussian}), the field magnitude ratio $R_{LT}$ for vortex beams is independent of the beam waist in the focal plane, and, even more surprisingly, independent of the propagation distance $z$, if the radial position is much smaller than the beam waist $\rho \ll~ w(z)$. The spatial distribution of the field ratio $R_{LT}$ is invariant under beam diffraction; it is constant along the entire unbounded axis of the beam. The geometrical surfaces where the ratio $R_{LT}$ is constant are cylinders of fixed radius around the infinite length of the beam optical vortex. This analytical result obtained from the paraxial approximation is verified in full-vectorial non-paraxial 3D field simulations in Section 4
and persists even under strong focusing conditions.
It should be noted that Eqs.~(\ref{eq:Eperp},\ref{eq:paraxial}) for the case of anti-aligned spin and orbital angular momenta  $(\sigma\cdot l<0)$ result in $E_z\propto\rho^{l-1}$ while $E_\perp\propto\rho^l$ and, hence, the dominance of the longitudinal field component in the vicinity of the optical vortex axis. We also note a $\pi/2$ phase shift between the longitudinal and transverse field components for linear polarisation, as follows from the Maxwell's equations combined with a paraxiality condition.

\section{Polarisation of 3D Vortex Fields}
An arbitrary complex 3D vector field $\mathbf E (x,y,z)$, can be expanded in terms of unit vectors $\hat{\mathbf n}$ $(n=x,y,z)$ in a Cartesian basis as 
$\mathbf E=~\sum_n E_n \hat{\mathbf n}$. The same field can be represented in a helicity basis as $\mathbf E = \sum_{\pm,z} E^{\pm,z} \hat{\mathbf{e}}_{\pm,z}$,
where $E^\pm= \frac{1}{\sqrt{2}} (\mp E_x+i E_y)$, and $\hat{\mathbf{e}}_{\pm}= \frac{1}{\sqrt{2}} (\mp \hat{\mathbf{x}}-i \hat{\mathbf{y}})$, $\hat{\mathbf{e}}_{z} = \hat{\mathbf{z}}$.
The polarisation coherence matrix for electric optical fields, $ E_m E^*_n$ (an asterisk indicates a complex conjugation), is fully defined in terms of standard Stokes parameters $S_{0-3}$ (see the Appendix) only if the longitudinal component of the field $E_z$ is neglected. 
However, as shown in the previous section, the longitudinal field of optical vortices is not negligible and may even be dominant at certain regions across the wavefront. For this reason, a Stokes description becomes incomplete and the formalism for field polarisation has to include all three components of the field. The matrix $\rho_{mn}=E_m E^*_n/|\mathbf{E}|^2$
is Hermitian by construction and can be fully defined by nine real parameters.   
%For description of optical polarisation of 3D fields, the parameters can be chosen to be coefficients of expansion in terms of Gell-Mann matrices \cite{carozzi2000}; its geometrical (polarisation ellipsoid) interpretation was discussed in Ref.~\cite{Dennis_2004}.  
Here, we will follow the convention  previously adopted for  description of polarisation of spin-1 particles \cite{Ohlsen_1972,Afanasev_2020}:
\begin{equation}
\rho_{mn}=\frac{1}{3}\Big\{ I+\frac{3}{2}\sum_{i=x,y,z} p_i \mathcal P_i+ 
\sum_{i,j=x,y,z} p_{ij}\mathcal P_{ij}\Big\}_{mn}, 
\end{equation}
where $I_{mn}$ is an identity matrix, and $(\mathcal{P}_i)_{mn}$ and  $(\mathcal{P}_{ij})_{mn}$ are the matrices of the spin vector and the quadrupolar tensor. The choice of normalisation $\sum_m \rho_{mm}=1$ reduces the number of independent parameters to eight. For a comprehensive treatise, see also Ref.~\cite{varshalovich1988quantum}, noting that the definition of $p_{nm}$ used here have an extra factor of 3 compared to Ref.~\cite{varshalovich1988quantum}.

The  corresponding vector and quadrupole polarisation parameters, $p_i$ and $p_{ij}$, can be expressed in terms of the field amplitudes $E_n$ and $E^{\pm,z}$ as
\begin{align}\label{eq:poldefs}
    \begin{split}
        |\mathbf{E}|^2p_n&=i \sum_{jk} \epsilon_{njk}E_j E_k^*,\\
        |\mathbf{E}|^2p_{nk}&=-\frac{3}{2} (E_n E_k^*+E_k E_n^*-\frac{2|\mathbf{E}|^2}{3}\delta_{nk}),
    \end{split}
\end{align}
leading to the following formulae for the independent polarisation parameters:
\begin{align}\label{eq:qpol}
    \begin{split}
        &|\mathbf{E}|^2p_x=\frac{1}{\sqrt 2} ((E^-+E^+)E_z^{*}+E_z(E^-+E^+)^*), \\
    &|\mathbf{E}|^2p_y=\frac{i}{\sqrt 2} (E_z(E^--E^+)^*-(E^--E^+)E_z^{*}), \\
    &|\mathbf{E}|^2p_{z}=|E^+|^2-|E^-|^2, \\
    &|\mathbf{E}|^2(p_{xx}-p_{yy})=3 (E^+E^{-*}+E^-E^{+*}),\\ 
    &|\mathbf{E}|^2p_{zz}=|E^+|^2+|E^-|^2-2|E_z|^2,  \\
    &|\mathbf{E}|^2p_{xy}=i\frac{3}{2}(E^+ E^{-*}-E^-E^{+*}),\\
    &|\mathbf{E}|^2p_{xz}=\frac{3}{2\sqrt{2}}(E^+E_z^{*}+E_zE^{+*}-E^-E_z{*}-E_zE^{-*}),  \\
    &|\mathbf{E}|^2p_{yz}=i\frac{3}{2\sqrt{2}}(E^+E_z^{*}-E_zE^{+*}+E^-E_z^{*}-E_zE^{-*}). 
    \end{split}
\end{align}
It follows from Eq.~(\ref{eq:qpol}) that the polarisation parameters have the following bounds:  $-3\leq (p_{xx}-p_{yy}) \leq 3$, $-2\leq p_{nn} \leq 1$, $-\frac{3}{2}\leq p_{nm}\leq \frac{3}{2}$ ($n\neq m$), and $-1\leq p_n\leq1$. The tensor of quadrupole polarisation is symmetric and traceless: $p_{nm}=p_{mn}$, $\sum_n p_{nn}=0$. 
In atomic and nuclear physics, these quantities are commonly referred to as orientation ($p_n$) and alignment ($p_{nm}$). 
%These polarisation parameters are in one-to-one correspondence to polarisations of localized atoms excited by this field via electric-dipole transitions \cite{Afanasev_2020}. 
Vector polarisation $p_n$ is crucial for describing properties of photonic skyrmions \cite{du2019deep}; its transverse components in evanescent fields were recently studied in Ref.~\cite{eismann2021transverse}.

In a limiting case of plane waves propagating in the $z$-~direction, the above defined polarisation parameters either turn to zero or reduce to Stokes parameters (Ref.\cite{collett2005field} and  the Appendix):
\begin{equation}\label{eq:stokeparameters}
 p_z\to \frac{S_3}{S_0},\ p_{xx}-p_{yy}\to -\frac{3S_1}{S_0},\ p_{xy}=-\frac{3S_2}{2S_0}, 
\end{equation}
while $p_{zz}\to 1$, indicating that the electric field of plane waves is transverse with respect to the $z$-axis. Another convention for the description of optical polarisation in 3D fields uses an expansion in terms of Gell-Mann matrices \cite{carozzi2000}, and is equivalent to the approach presented here.

The ratio $R_{LT}$ introduced in a previous section may be probed experimentally by measuring the polarisation parameter $p_{zz}$. 
The independence of $R_{LT}$ from the beam waist found in the previous section has immediate implications for $p_{zz}$ which, as a result, maintains constant transverse spatial dimensions independently of beam divergence due to diffraction. Using Eq.~(\ref{eq:Rltcases}) and the definitions from Eq.~(\ref{eq:qpol}), $z$-independent expressions can be obtained for $p_{zz}$:
\begin{eqnarray}
\label{eq:pzz}
p_{zz}(\zeta)= 
\begin{cases}
\frac{1-4\zeta^2}{1+2\zeta^2},\ \ \text{circular}\  (\sigma\cdot l< 0) \ \text{polarisation}\\
1,\ \ \ \ \ \ \ \ \ \ \text{circular}\ (\sigma\cdot l>0)\ \text{polarisation}\\
\frac{1-2\zeta^2}{1+\zeta^2},\ \ \ \text{linear polarisation}\\
\frac{1-8\zeta^2}{1+4\zeta^2},\ \ \  \text{radial polarisation}\\
1,\ \ \ \ \ \ \ \ \ \ \  \text{azimuthal polarisation}\\ \,.
\end{cases} 
\end{eqnarray}
For a circularly polarised beam with $\sigma\cdot l< 0$, it follows from Eq.~(\ref{eq:pzz}) that $p_{zz}=-2$ in the vortex center and $p_{zz}$ approaches unity with increasing radial distance to the singularity. Zero crossing ($p_{zz}=0$) takes place at $\rho=|l|\lambda/\pi$ and is independent of both the beam waist and the propagation distance $z$ (see insets of Fig.~\ref{fig:linpolvortex}(c)), while increasing linearly with $l$.
%It follows that the transverse dimensions of polarisation structures increase linearly with the topological charge $l$.
%For linear polarisation, zero crossing take place at $\rho=\frac{|l|\lambda}{\pi \sqrt{2}}$. 
%Choosing $l=1$, we observe that $p_{zz}$ varies from its minimum value of $-2$ to zero over the radial distance of $\frac{\rho}{\lambda}=\frac{1}{\pi}\approx 0.32$ for circular and $\frac{\rho}{\lambda}=\frac{1}{\pi\sqrt{2}}\approx 0.23$ for linear polarisation, respectively. This spatial variation at a sub-wavelength scale remains unchanged for arbitrarily long distances from beam's focus.
Similar propagation-independent expressions may be obtained for the other polarisation parameters in Eq.~(\ref{eq:qpol}), with $\zeta$ playing the role of a scaling variable (see the Appendix for details).
%, $i.e.$ its angular position at $z\to\infty$ reduces to zero.
%This property of polarization is due to non-diffractive behavior of the ratio $R_{LT}$. 

The above results were obtained in a simplified analytical model for a \textit{paraxial} optical vortex field. Next, we demonstrate the non-diffractive behaviour of these polarisation features using numerical simulations for a \textit{non-paraxial} field.

\section{Numerical Approach and Discussion}

\begin{figure}[b!]
    \centering
    \includegraphics[]{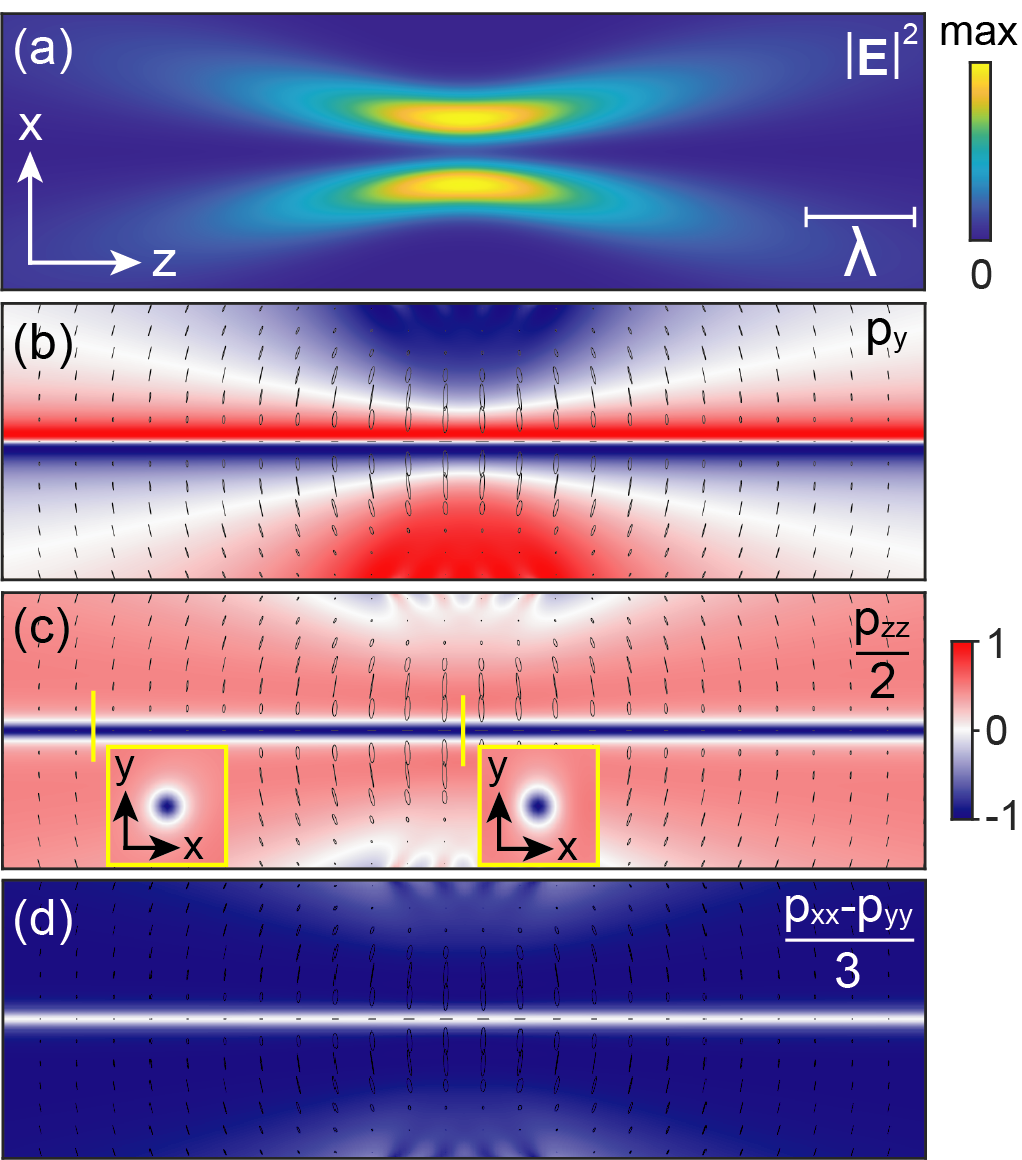}
    \caption{Polarisation parameters for a focused Laguerre-Gaussian vortex beam with $l=1$, linearly polarised along the $x$-direction and propagating along the $z$-axis. The beam waist is $w_0=\lambda$.
    (a) The intensity distribution of the beam in the ($xz$)-plane. (b-d) Colourmaps of the $p_y$, $p_{zz}$ and $p_{xx}-p_{yy}$ polarisation parameters with polarisation ellipses overlaid on top, indicating the polarisation state at each point in space. The polarisation structure around the phase singularity is non-diffractive and invariant with respect to the beam waist. Some parameters are scaled to fit within the $\pm1$ colour scale range.}
    \label{fig:linpolvortex}
\end{figure}

%Our approach for calculating the full 3D electromagnetic fields of a focused beam is described in detail in Ref. \cite{JKSarxiv}. 
% reference TBD; paper not submitted yet
We now outline a full-wave, non-paraxial, numerical approach. Any monochromatic electromagnetic field can be decomposed into a spectrum of plane wave components with wavevectors $\mathbf{k} = k_x \hat{\mathbf{x}} + k_y \hat{\mathbf{y}} + k_z \hat{\mathbf{z}}$ lying on the k-sphere of radius $k = \frac{\omega}{c}$, and hence $k_z~=~\sqrt{k^2-k_x^2-k_y^2}$, as follows:

\begin{equation}\label{eq:integrateangularspectrum}
    \mathbf{E}(\mathbf r) = \iint (\mathcal{A}_{p} \hat{\mathbf{e}}_p + \mathcal{A}_{s} \hat{\mathbf{e}}_s) \, e^{i(\mathbf{k} \cdot \mathbf{r})} \text{d} k_x \, \text{d} k_y,
\end{equation}

\noindent where $\mathcal{A}_{p/s}(k_x,k_y)$ are the components of the angular spectrum pertaining to each of the orthonormal polarisation basis vectors, which we take as $\hat{\mathbf{e}}_s= \frac{1}{\sqrt{k_x^2+k_y^2}} \, (-k_y \, \hat{\mathbf{x}} + k_x \, \hat{\mathbf{y}})$ and $\hat{\mathbf{e}}_p=\hat{\mathbf{e}}_s \times \frac{\mathbf{k}}{k}$, corresponding to the azimuthal and polar angle spherical basis vectors tangential to the k-sphere \cite{Lakhtakia1992,Rotenberg2012,Picardi2017}. Equation (\ref{eq:integrateangularspectrum}) is an integration of plane waves and constitutes an exact solution to Maxwell's equations, including all the field components. In order to compute it one only needs to find the spectral amplitudes $\mathcal{A}_{p/s}(k_x,k_y)$ corresponding to the desired illumination. 

For these exact field calculations, we employ the Laguerre-Gaussian vortex beam widely used in singular optics \cite{Saleh2007}. The angular spectrum of a Laguerre-Gaussian vortex beam can be calculated by selecting the $z=0$ plane and performing a Fourier transform $(x,y)\to(k_x,k_y)$ \cite{Picardi2017,Devaney1974}. The derivation of the plane wave polarisation amplitudes $\mathcal{A}_{p/s}(k_x,k_y)$ from the paraxial Laguerre-Gaussian beam field is described in detail in the Appendix.

We can use Eq.~(\ref{eq:integrateangularspectrum}) to calculate the full 3D electric field and plot required polarisation parameters without any approximations. The first example is a linearly polarised non-paraxial vortex beam with $l=1$ and $w_0 = \lambda$ propagating along the positive $z$-axis (Fig.~\ref{fig:linpolvortex}). The tight focusing creates a highly divergent beam. In this case, the $p_y$, $p_{zz}$ and $p_{xx}-p_{yy}$ parameters are required to fully describe the polarisation structure (please note that the only remaining nonzero polarisation parameter is $p_{xz}$ but its behaviour in the $(yz)$-plane is the same as that of $p_y$ in the $(xz)$-plane; see Fig.~\ref{fig:linpolvortex}(b)). In contrast to the electric field, which diffracts naturally after a propagation distance of just a few wavelengths (Fig.~\ref{fig:linpolvortex}(a)), all three polarisation parameters in Fig.~\ref{fig:linpolvortex}(b-d) clearly show  no divergence around the phase singularity that lies on the $z$-axis. The polarisation structures remain invariant and extend far beyond the focal plane, in agreement with the analytical paraxial predictions for $R_{LT}$ in Eq.~(\ref{eq:Rltcases}), but numerically observed here beyond the paraxial approximation.

\begin{figure}[t]
    \centering
    \includegraphics[width=13cm]{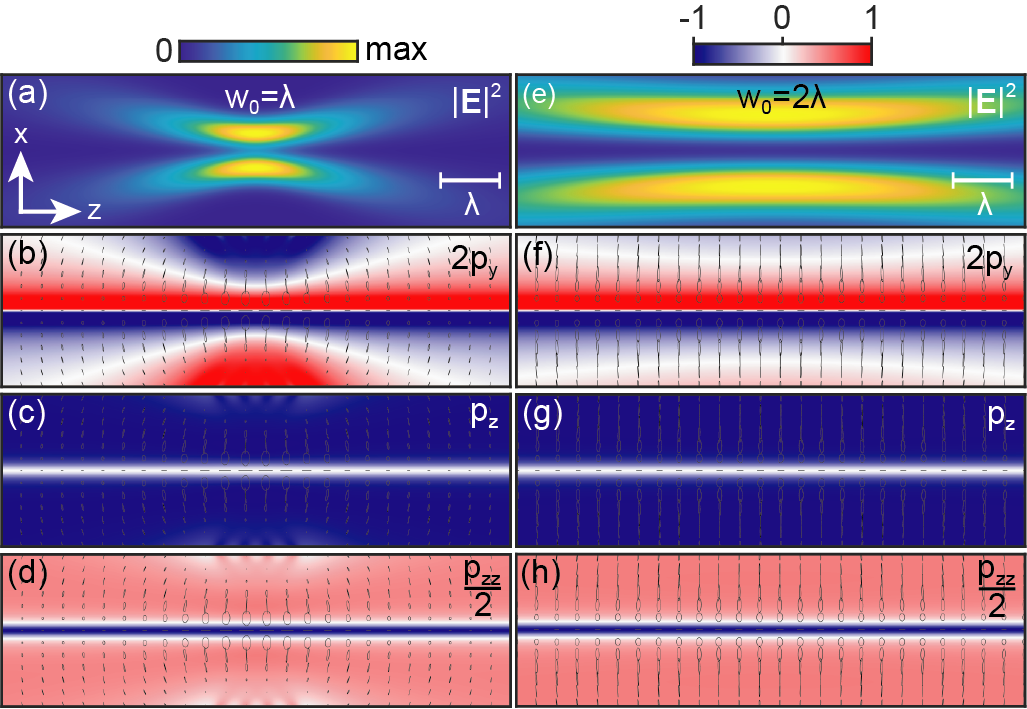}
    \caption{Polarisation parameters for a focused Laguerre-Gaussian vortex beam with $l=1$ and $\sigma=-1$ (left-hand circularly polarised), propagating along the positive $z$-axis. The beam waist is (a-d) $w_0=\lambda$ and (e-h) $w_0=2\lambda$. 
    (a \& e) The intensity distribution of the beam in the ($xz$)-plane. (b-d \& f-h) Colourmaps of the $p_y$, $p_z$ and $p_{zz}$ polarisation parameters with polarisation ellipses overlaid on top, indicating the polarisation state at each point in space. The polarisation structure around the phase singularity is non-diffractive and invariant with respect to the beam waist. Some parameters are scaled to fit within the $\pm1$ range of color scale.}
    \label{fig:circpolvortex}
\end{figure}

We now demonstrate the equivalent polarisation properties of a circularly polarised vortex beam. Figure \ref{fig:circpolvortex}(a) shows the intensity of a circularly polarised vortex beam, similar in many respects to the previous case except with $\sigma = -1$, again calculated using Eq.~(\ref{eq:integrateangularspectrum}). Note that the spin and orbital angular momenta are anti-aligned in this beam. The beam waist is set to $w_0 = \lambda$ so the beam electromagnetic field is highly focused  and diffraction is significant. The polarisation parameters in Eq.~(\ref{eq:qpol}) are plotted in Fig.~\ref{fig:circpolvortex}b-d. All three plots show a non-divergent polarisation structure around the beam axis.  Figure \ref{fig:circpolvortex}(e-h) shows the same results for a vortex beam with a beam waist of double the size and, therefore, weaker focusing. Nevertheless, the non-diffractive polarisation structure near the phase singularity is unperturbed, with the $p_{zz}=0$ (white) contour lying at $\rho = 0.32 \lambda $ in both Figs.~\ref{fig:circpolvortex}(d) and \ref{fig:circpolvortex}(h), in good agreement with the analytical result from Eq.~(\ref{eq:pzz}): $\rho=|l|\lambda/\pi$. This structural invariance has been observed for all beam waists, independent of focusing.

\begin{figure}[b]
    \centering
    \includegraphics[]{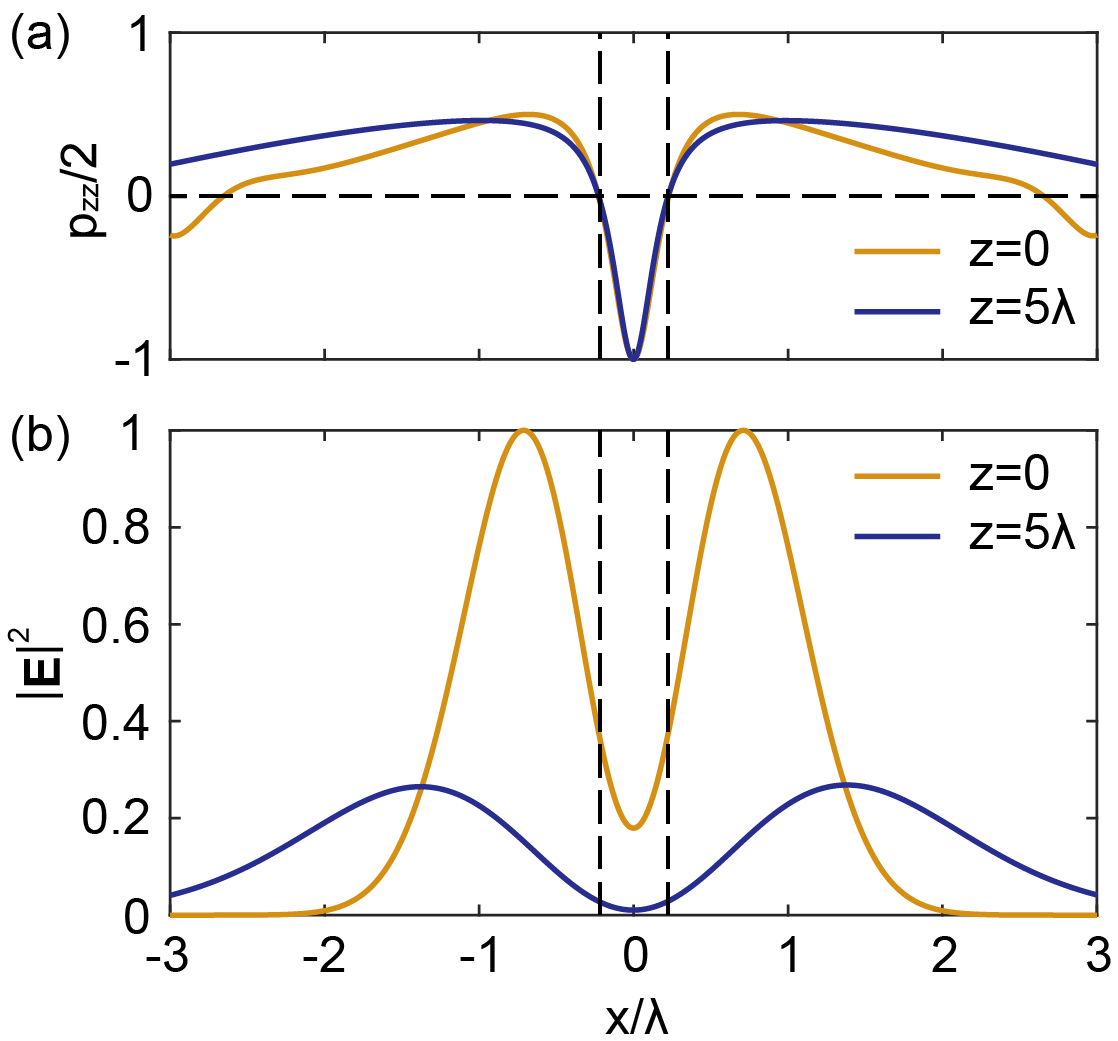}
    \caption{Cross-sections of the linearly polarised vortex beam with $w_0=\lambda$ depicted in Fig.~\ref{fig:linpolvortex} showing (a) $p_{zz}$ and (b) $|\mathbf{E}|^2$ at the focal plane and after propagating a distance of $5\lambda$. The black dotted guidelines show the points at which $p_{zz}=0$, and the corresponding field intensity for each cross-section plane.}
    \label{fig:crosssectionpzz}
\end{figure}

The main challenge in detecting this non-diffractive polarization property is the requirement to perform measurements in a region of space where the field intensity is weaker compared to its maximum. The ability to detect it is determined by the sensitivity of the detection apparatus, and can be mediated to some degree by the choice of wavelength, beam waist, how far the detection plane is from the focal plane and what polarisation structure is being investigated. 
%Fig.~\ref{fig:crosssectionpzz} is provided to demonstrate that this phenomenon can in principle be detected over short propagation distances but the challenge lies with detection over large distances. 
Figure \ref{fig:crosssectionpzz}(a) shows the cross-sectional plots of the $p_{zz}$ parameter for the linearly polarised $l=1$ vortex beam in Fig.~\ref{fig:linpolvortex} at different points along the $z$-axis. As before, we see non-diffractive behaviour when $p_{zz}=0$ near the beam's vortex, indicated by a horizontal dotted black line. Note how other features where $p_{zz}=0$ at locations further away from the beam centre are subjected to diffraction. The intensity of the normalised electric field in the same cross-section planes at the location of the non-diffracting polarizatio features, indicated by the vertical dotted black lines, is 37\% of the peak intensity at the focal plane, and $2.8\%$ at $z=5\lambda$ (Fig.~\ref{fig:crosssectionpzz}(b)). The $p_{zz}=0$ polarisation structure should, therefore, be easily detectable in the focal plane of the beam and measurable away from the focus. The intensity drop-off of a beam is dictated by the Rayleigh range which is proportional to the square of the beam waist. However, increasing the beam waist reduces the intensity of the longitudinal field. The resulting optimisation will depend on the measurement sensitivity and the desired application. 

%In the interest of future experimental observations, we repeated the simulations in a model that accounts for the beam diffraction due to focusing by a lens with a finite numerical aperture (NA). 
%This was done by restricting the integration of the ($k_x,k_y$)-plane in Eq. (\ref{eq:integrateangularspectrum}) down from $k_x^2+k_y^2 \leq k^2$ to $k_x^2+k_y^2 \leq (\text{NA})^2 k^2$. This approach, therefore, filters out the largest transverse wavevector components, and replicates one of the primary effects of a finite size lens (see the Appendix for details). The results show that the fields close to the beam axis are resilient with respect to a finite NA, and only relatively far from the beam axis do additional fringes in the beam profile arise. Once the NA is small enough to significantly crop the angular spectrum of the beam, the beam is simply augmented to resemble a less tightly focused beam, but the non-diffractive polarisation parameters persist. Numerical introduction of the defects in the propagation path and astigmatism have also no effect on the described polarisation feature (Fig. \ref{fig:NAlim}). We therefore conclude that the non-diffractive polarisation structures in this work should be observable experimentally. 

\begin{figure}[b]
    \centering
    \includegraphics[width=\linewidth]{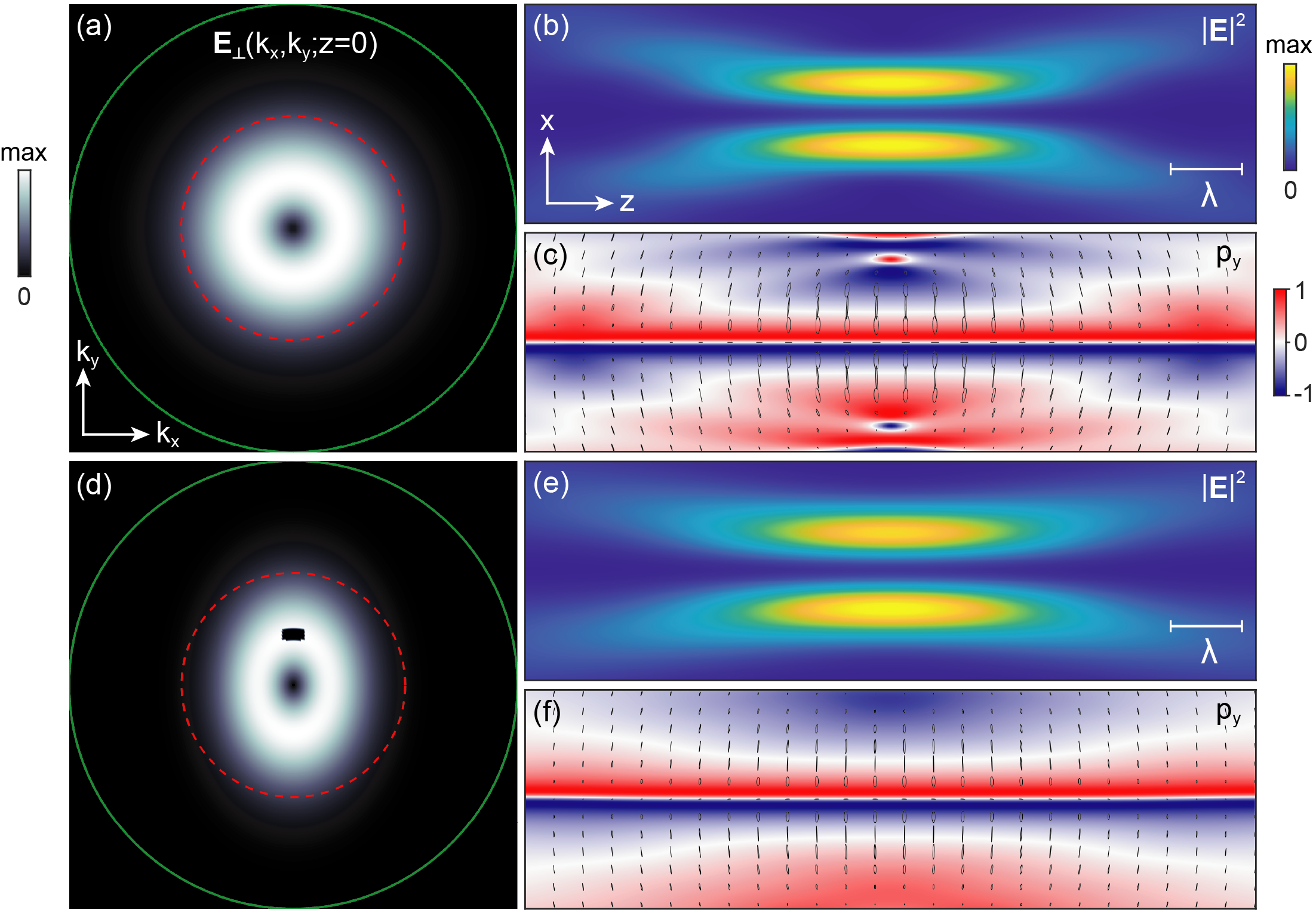}
    \caption{(a) The angular spectrum of a linearly polarised ($x$-direction) vortex beam with $l=1$. The green line indicates the light line and the red dotted line indicates the inner limit of the cropped region dictated by NA $=0.5$. (b) The intensity distribution of the corresponding vortex beam with NA $=0.5$ generated by integrating the field distribution in (a). The beam waist is $w_0=\lambda$. (c) Colourmap of the $p_y$ polarisation parameter with polarisation ellipses overlaid on top, indicating the polarisation state at each point in space. The non-diffractive behaviour near the phase singularity is maintained and only peripheral fields are affected by the numerical aperture reduction. (d-f) The same quantities as in (a-c) but with part of the angular spectrum removed and astigmatism applied along $x$.}
    \label{fig:NAlim}
\end{figure}

In order to experimentally verify the propagation-invariant polarisation structures, a vortex beam will likely need to be focused using a lens with a defined numerical aperture (NA). The effect of a restricted NA was simulated by limiting the integration of the ($k_x,k_y$)-plane in Eq.~(\ref{eq:integrateangularspectrum}) from $k_x^2+k_y^2 \leq k^2$ to $k_x^2+k_y^2 \leq (\text{NA})^2 k^2$. Figure \ref{fig:NAlim}(a) shows the non-zero component of the angular spectrum for a linearly polarised vortex beam with $l=1$ and $w_0=\lambda$, given by Eq.~(\ref{eq:lgangularspectrum}). This is equivalent to the back focal plane image of the beam and the phase singularity is clearly visible at $k_x=k_y=0$. In the main text, all fields within the light line (indicated by a green line) are integrated with Eq.~(\ref{eq:integrateangularspectrum}) to create the real-space field distribution. We crop the angular spectrum down to a factor of NA$\cdot k$ with NA $=0.5$ was used here, as indicated by the red dotted line. Figure \ref{fig:NAlim}(b) shows the intensity profile of the beam after this NA restriction. When compared with the ideal beam in Fig.~\ref{fig:linpolvortex}(b), the polarisation parameter of the restricted beam in Fig.~\ref{fig:NAlim}(c) reveals the same non-diffractive property near the phase singularity and only disturbances in the peripheral fields are observable. As the NA is reduced further (not shown), the beam waist widens but the behaviour around the phase singularity is maintained. 

One can proceed to add more imperfections or aberrations to the beam. A defect or a piece of dust on the focusing lens can perturb the beam. This can be approximated by deleting part of the angular spectrum. A lens can also introduce astigmatism to the beam. The effect can be roughly modelled by scaling $k_x$ and $k_y$ in the angular spectrum. Fig.~\ref{fig:NAlim}(d) shows the angular spectrum of the same vortex beam as Fig. \ref{fig:NAlim}(a) with a restricted NA of 0.5 but with these two previously mentioned additional perturbations applied. The angular spectrum is set to zero for $-0.05<k_x<0.05$ and $0.2<k_y<0.25$, and $k_x$ is transformed by $k_x \to 0.75 k_x$, therefore, reciprocally stretching the beam in the $x$-direction. A non-diffractive nature of the $p_y$ polarisaton parameter near the phase singularity is preserved for such scattered focused beams with astigmatism (Fig.~\ref{fig:NAlim}(f)). We therefore conclude that the non-diffractive polarisation structures within a vortex beam should be robust to a variety of experimental imperfections and experimentally observable in this respect.

\section{Conclusions}
We have studied the polarisation of vector beams carrying optical angular momentum. We show the existence of polarisation features within optical vortex beams which maintain constant transverse spatial dimensions independently of the beam divergence due to diffraction. The exact size of these vortex polarisation structures is dictated by the presence of the longitudinal electric field in the beam, and such structures are expected for vortex beams of all topological charges. An analytical paraxial model predicts their presence in weakly focused beams and a numerical angular spectrum approach further extended this prediction to tightly focused beams, thereby proving applicability to all vortex beams. These polarisation features are not affected by finite numerical apertures and so should be experimentally measurable. It should be noted that the predicted non-diffractive polarisation features have relatively small transverse dimensions of the order $\Delta\rho\approx l\lambda/\pi$, centered on a low-intensity region of the optical vortex wavefront. Therefore, future measurements will require sub-wavelength resolution at low $l$ and increased sensitivity of the probe for larger values of $l$.

The demonstrated effect allows one to pinpoint the position of a phase singularity with subwavelength accuracy independently of the size of a beam spot. This property may have useful applications in metrology, optical communications, optical networking, laser sensing and radar operations.

%%%%%%%%%%%%%%%%%%%%%%%%%%%%%%%%%%
\paragraph{Funding.}  This work was supported in part by the ERC iCOMM Project (No. 789340) and the ERC Starting Grant No. ERC-2016-STG-714151-PSINFONI. Work of A.A. was supported by the US ARO under Grant W911NF-19-1-0022.

\paragraph{Disclosures.} The authors declare that there are no conflicts of interest related to this article.

\paragraph{Data availability.} All the data supporting finding of this work are presented in the Results section and are available from the corresponding author upon reasonable request.
%No data were generated or analyzed in the presented research.

%\paragraph{Supplemental Information.} See Supplement for supporting content.

%\paragraph{Acknowledgements.}

%%%%%%%%%%%%%%%%%%%%%%%%%%%%%%%%%%

\bibliography{vortex}

%\bibliographyfullrefs{vortex}

%\newpage

\setcounter{figure}{0}
\setcounter{equation}{0}
\setcounter{section}{0}
\renewcommand{\thefigure}{A\arabic{figure}}
\renewcommand{\theequation}{A\arabic{equation}} 

%\begin{center}
%	\textbf{APPENDIX} 
%\end{center}

\begin{center}
\section*{Appendix}
\end{center}

\section{Calculation of Polarisation Parameters}
Here we present analytic expressions for polarisation parameters (Eq.~(\ref{eq:qpol})) calculated for different polarisations of optical vortex beams; $l>0$ is assumed. The transverse field is defined by Eq.~(\ref{eq:Eperp}) and the longitudinal field is obtained from $\mathbf \nabla \cdot \mathbf E=0$ combined with the paraxiality condition, Eq.~(\ref{eq:paraxial}), at radial positions $\rho$ near the beam's axis much smaller than the beam waist. As in the main text, we define the dimensionless radial parameter  $\zeta \equiv l/(k \rho)$ which depends on the topological order parameter $l$. The transverse polarisation vector \text{\boldmath$\eta_\perp$} for different polarisations is given in terms of unit vectors in Cartesian or cylindrical coordinates.

\begin{center}
\begin{tabular}{ |l|c|c|c|c|c| }
\hline
 & Circular  & Circular  & Linear & Radial & Azimuthal\\
 & $\sigma\cdot l<0$ & $\sigma\cdot l>0$ &  & &\\
 \hline
 \text{\boldmath$\eta_\perp$} & $\frac{1}{\sqrt{2}}(\hat{\mathbf x}-i\hat{ \mathbf y})$ & $-\frac{1}{\sqrt{2}}(\hat{\mathbf x}+i\hat{\mathbf y})$ & $\hat{\mathbf x}$ & $\hat{\boldsymbol{\rho}}$ & $\hat{\boldsymbol{\phi}}$\\ 

\hline
 & & & & &\\
 $p_z$ & $-\frac{1}{1+2\zeta^2}$ & 1 & 0 & 0 & 0\\ 
  & & & & &\\
\hline

 & & & & &\\
 $p_y$ & $\frac{\sqrt{2}|\zeta|\cos{\phi} }{1+2\zeta^2}$  & 0 & $-\frac{2|\zeta|\cos{\phi}}{1+\zeta^2}$ &$-\frac{4|\zeta|\cos{\phi}}{1+4\zeta^2}$   & 0\\
  & & & & &\\
 \hline
 
  & & & & &\\
 $p_x$ & $-\frac{\sqrt{2}|\zeta|\sin{\phi} }{1+2\zeta^2}$ & 0 & 0 & $\frac{4|\zeta|\sin{\phi}}{1+4\zeta^2}$  &0\\
  & & & & &\\
  \hline
  
    & & & & &\\
 $p_{xx}-p_{yy}$ & 0 & 0 & $\frac{3}{1+\zeta^2}$ & $-\frac{3\cos{2\phi}}{1+4\zeta^2}$ & $3\cos{2\phi}$\\
  & & & & &\\
  \hline
  
      & & & & &\\
 $p_{xy}$ & 0 & 0 & 0 & $-\frac{3}{2}\frac{\sin{2\phi}}{1+4\zeta^2}$ &$\frac{3}{2}\sin{2\phi}$\\
  & & & & &\\
  \hline
     & & & & &\\
 $p_{xz}$ & $-\frac{3}{\sqrt{2}}\frac{|\zeta|\sin{\phi}} {1+2\zeta^2}$  & 0 & $\frac{3|\zeta|\sin{\phi}}{1+\zeta^2}$ & 0 & 0\\
  & & & & &\\
  \hline 
      & & & & &\\
 $p_{yz}$& $\frac{3}{\sqrt{2}}\frac{|\zeta|\cos{\phi}} {1+2\zeta^2}$ & 0 & 0 & 0 &0\\
  & & & & & \\
  \hline
        & & & & &\\
 $p_{zz}$& $\frac{1-4\zeta^2}{1+2\zeta^2}$ & 1 & $\frac{1-2\zeta^2}{1+\zeta^2}$ & $\frac{1-8\zeta^2}{1+4\zeta^2}$ & 1\\
 & & & & &\\
  \hline
\end{tabular}
\end{center}
The expressions for circular polarisation ($\sigma\cdot l<0$) are simplified in cylindrical coordinates, for which radial components of the polarisation vector and tensor are zero: $p_\rho=0,\ p_{\rho z}=0,\ p_\phi=\frac{\sqrt{2}|\zeta|}{1+2\zeta^2}, p_{\phi z}=\frac{3}{\sqrt{2}}\frac{|\zeta|} {1+2\zeta^2}$.

\section{Definitions of Stokes Parameters}

A polarisation coherence matrix $E_m E_n^*$ for 2D electric fields $\mathbf E=(E_x,E_y,0)$ is defined in terms of Stokes parameters as \cite{collett2005field}:
\begin{equation}
  S_0=E_x E_x^*+E_y E_y^*,\
  S_1=E_x E_x^*- E_y E_y^* ,\
  S_2=E_x E_y^*+E_y E_x^*,\ 
  S_3=i (E_x E_y^*-E_y E_x^*). 
\end{equation}
If the transverse electric field is linearly polarised along $x$-axis, then the polarisation of the full field is determined by three Stokes parameters defined for its corresponding $x,z$-components which, in turn, relate to the polarisation parameters in Eq.~(\ref{eq:qpol}) as: 
\begin{equation}\label{eq:stokesXZ}
  \frac{S_1^{(xz)}}{S_0}=\frac{p_{zz}-p_{xx}}{3},\ 
  \frac{S_2^{(xz)}}{S_0}=-\frac{2}{3}p_{xz}, \
  \frac{S_3^{(xz)}}{S_0}=-p_y. 
\end{equation}

%\section{Non-diffractive Polarisation behaviour in an analytic model}

 %Given the model transverse field in Eq. (\ref{eq:Eperp}), the beam intensity profile evolving with the propagation distance $z$ is shown in Fig. \ref{fig:an2D} (left) in comparison to the propagation-invariant $p_{zz}$ polarisation profile (Fig. \ref{fig:an2D} (right)) in the case of circular polarisation with $\sigma\cdot l<0$, $w_0=2\lambda,\ l=1$, and $\sigma=-1$. A slight change of polarisation at small $z$ is due to the fact that the beam waist $w_0$ was chosen to be not much larger than the wavelength.
 
% \begin{figure}[h]
%    \centering
%    \includegraphics[width=6.cm]{Figures/anIntensity2D.png}
%    \includegraphics[width=6.cm]{Figures/anPzz2D.png}
%    \caption{(Left) The beam intensity profile for different propagation distances $z=0$ and $5z_R$ compared to (right) the $p_{zz}$ polarisation profile.}
%    \label{fig:an2D}
%\end{figure}

%\begin{figure}[h!]
%   \centering
%    \includegraphics[width=7.cm]{Figures/anInt3D_z=0.png}
 %   \includegraphics[width=7.cm]{Figures/anInt3D_z=5zR.png}
%   \includegraphics[width=7.cm]{Figures/anPzz3D.png} 
%   \caption{The beam intensity profile for (top left) $z$=0 and (top right) $z=5 z_R$ and (bottom) the propagation-independent polarisation profile. Note the difference in the transverse scales for intensities and polarisation.}
%    \label{fig:an3D}
%\end{figure}

\section{Decomposing the angular spectrum into a polarisation basis}

Here we show how non-paraxial fields of a focused vortex beam are calculated using the angular spectrum approach. 
%We explain how the amplitudes $\mathcal{A}_{p/s}$ used in Eq. (\ref{eq:integrateangularspectrum}) can be computed from Eq. (\ref{eq:Eperp}).
%We start with the transverse fields $\mathbf{E}_\perp(\mathbf{r})$ given in Eq. (\ref{eq:Eperp}) and perform a Fourier transform in the $z=0$ plane:
We start with the paraxial expression for a Laguerre-Gauss beam \cite{Saleh2007}, 
\begin{equation}\label{eq:fullLGbeam}
    \mathbf{E}_\perp(\mathbf{r}) = \text{\boldmath$\eta_\perp$} A_\perp \frac{w_0}{w(z)} \bigg(\frac{\rho}{w(z)}\bigg)^{|l|} \mathcal{L}_m^l\bigg(\frac{2 \rho^2}{w^2(z)}\bigg) e^{-\frac{\rho^2}{w^2(z)}} e^{i(l\phi + kz + k \frac{\rho^2}{2R(z)} -(l+2m+1)\zeta(z))},
\end{equation}
where $w_0$ is the beam waist in the focal plane, $w(z)$ is the beam radius at any point in space, $\mathcal{L}_m^l$ is the generalised Laguerre polynomial of order $l$ and a radial index $m$, $R(z)$ is the beam curvature radius, and $\zeta(z)$ is the Gouy phase factor. 
We then consider a Fourier transform on the $z=0$ plane which defines the angular spectrum of a beam,
\begin{equation}
    \mathbf{E}_\perp(k_x,k_y) = \frac{1}{4 \pi^2} \iint \mathbf{E}_\perp(x,y) \, e^{-i (k_x x + k_y y)} \, \text{d}x \, \text{d}y.
\end{equation}
This Fourier transform can be solved analytically. For an $l=1$ and $m=0$ Laguerre-Gauss beam,
\begin{equation}\label{eq:lgangularspectrum}
    \mathbf{E}_\perp(k_x,k_y) = i \text{\boldmath$\eta_\perp$} A_\perp \frac{\pi w_0^3}{2} e^{\frac{-w_0^2(k_x^2+k_y^2)}{4}} (k_x + i k_y). 
\end{equation}
This is the angular spectrum of the transverse components only (it ignores the $z$-component), but from $\mathbf{E}_\perp(k_x,k_y)$  one can find the plane wave amplitudes $\mathcal{A}_{p/s}(k_x,k_y)$ that, when substituted into Eq.~(\ref{eq:integrateangularspectrum}), give an electric field $\mathbf{E} = \mathbf{E}_\perp + E_z \mathbf{\hat{e}}_z$ at $z=0$ (the focal plane), whose transverse component matches exactly Eq.~(\ref{eq:fullLGbeam}), but which also possesses the corresponding $E_z$ component that appears naturally from the electromagnetic plane wave polarisation superposition. 

In the remaining of this section, we do not explicitly write the $(k_x,k_y)$ dependencies of the angular spectrum for ease of notation, but note that all the fields mentioned here are the spectra defined in the $(k_x,k_y)$-plane unless otherwise stated. All fields are assumed to be time-harmonic.

%We need to find the relation between the transverse fields $\mathbf{E}_\perp$ and the polarization spectra $\mathcal{A}_{p}, \mathcal{A}_{s}$. 
The angular spectrum of the total field $\mathbf{E}$ can be represented in a Cartesian basis $\mathbf{E} = E_x \hat{\mathbf{x}} + E_y \hat{\mathbf{y}} + E_z \hat{\mathbf{z}}$, which can be split into a transverse part and a longitudinal part $\mathbf{E} = \mathbf{E}_\perp + E_z \hat{\mathbf{z}}$. Similarly, in the \textit{p/s} polarisation basis $\mathbf{E} = \mathcal{A}_{p} \hat{\mathbf{e}}_p + \mathcal{A}_{s} \hat{\mathbf{e}}_s$. 
%\begin{align}
%    \mathbf{E} &= E_x \hat{\mathbf{x}} + E_y \hat{\mathbf{y}} + E_z \hat{\mathbf{z}}\\ 
%    &= \mathbf{E}_\perp + E_z \hat{\mathbf{z}} \\
%    &= \mathcal{A}_{p} \hat{\mathbf{e}}_p + \mathcal{A}_{s} \hat{\mathbf{e}}_s
%\end{align}
Equating these, one can write the transverse part of the field as
\begin{equation}\label{eq:EperpApAsEz}
    \mathbf{E}_\perp = \mathcal{A}_{p} \hat{\mathbf{e}}_p + \mathcal{A}_{s} \hat{\mathbf{e}}_s - E_z \hat{\mathbf{z}}.
\end{equation}
We can further write $E_z$ in terms of $\mathcal{A}_{p}$ as $E_z = \mathbf{E} \cdot \hat{\mathbf{z}} = (\mathcal{A}_{p} \hat{\mathbf{e}}_p + \mathcal{A}_{s} \hat{\mathbf{e}}_s) \cdot \hat{\mathbf{z}}= \mathcal{A}_{p} (\hat{\mathbf{e}}_p \cdot \hat{\mathbf{z}})$, where we used the fact that $\hat{\mathbf{e}}_s \cdot \hat{\mathbf{z}} = 0$ because $\hat{\mathbf{e}}_s= \frac{1}{\sqrt{k_x^2+k_y^2}} \, (-k_y \, \hat{\mathbf{x}} + k_x \, \hat{\mathbf{y}})$. Thus, the transverse field is uniquely related to the $s$ and $p$ polarisation amplitudes as
\begin{equation}\label{eq:EperpApAs}
    \mathbf{E}_\perp = \mathcal{A}_{p}\left[ \hat{\mathbf{e}}_p - (\hat{\mathbf{e}}_p \cdot \hat{\mathbf{z}}) \hat{\mathbf{z}}\right] + \mathcal{A}_{s} \hat{\mathbf{e}}_s.
\end{equation}
If we now perform a dot product with the p/s basis unit vectors, and noting that $\hat{\mathbf{e}}_p \cdot \hat{\mathbf{e}}_s = 0$ and $\hat{\mathbf{e}}_p \cdot \hat{\mathbf{e}}_p = \hat{\mathbf{e}}_s \cdot \hat{\mathbf{e}}_s = 1$, we find 
\begin{align}\label{eq:normalisedApAsdotepdotes}
    \begin{split}
        \mathbf{E}_\perp \cdot \hat{\mathbf{e}}_p &= \mathcal{A}_{p}\left( 1 - (\hat{\mathbf{e}}_p \cdot \hat{\mathbf{z}})^2\right), \\
        \mathbf{E}_\perp \cdot \hat{\mathbf{e}}_s &= \mathcal{A}_{s}. \\
    \end{split}
\end{align}
\noindent Knowing that 
$\hat{\mathbf{e}}_p=\hat{\mathbf{e}}_s \times \frac{\mathbf{k}}{k} = \Big(\frac{k_x k_z}{k \sqrt{k_x^2+k_y^2}}, \frac{k_y k_z}{k \sqrt{k_x^2+k_y^2}}, -\frac{\sqrt{k_x^2+k_y^2}}{k}\Big)$, we find that $\left( 1 - (\hat{\mathbf{e}}_p \cdot \hat{\mathbf{z}})^2\right) = \left(\frac{k_z}{k}\right)^2$, and so we can obtain the orthogonal scalar plane wave polarisation coefficients in terms of the transverse field spectrum as
\begin{align}\label{eq:normalisedApAs}
    \begin{split}
        \mathcal{A}_{p} &= (\mathbf{E}_\perp \cdot \hat{\mathbf{e}}_p) \bigg(\frac{k}{k_z}\bigg)^2\\
        \mathcal{A}_{s} &= \mathbf{E}_\perp \cdot \hat{\mathbf{e}}_s.
    \end{split}
\end{align}
Applied to the specific case of the vortex beam defined by Eq.~(\ref{eq:lgangularspectrum}), we obtain the explicit expressions for the \textit{p/s} polarisation basis coefficients,

\begin{align}
    \begin{split}
        \mathcal{A}_{p} &=  i A_\perp\frac{ w_0^3}{8 \pi} e^{\frac{-w_0^2(k_x^2+k_y^2)}{4}} (k_x+i k_y) \, k \frac{k_x \eta_x + k_y \eta_y}{k_z \sqrt{k_x^2+k_y^2}} \\
        \mathcal{A}_{s} &= i A_\perp\frac{ w_0^3}{8 \pi} e^{\frac{-w_0^2(k_x^2+k_y^2)}{4}} (k_x+i k_y) \frac{-k_y \eta_x + k_x \eta_y}{\sqrt{k_x^2+k_y^2}}, \\
    \end{split}
\end{align}
where $\text{\boldmath$\eta_\perp$} = \eta_x \hat{\mathbf{x}} + \eta_y \hat{\mathbf{y}}$. The non-paraxial fields are then generated by substituting the above $\mathcal{A}_{p/s}$ into Eq.~(\ref{eq:integrateangularspectrum}) from the main text. We numerically compute the integral in Eq.~(\ref{eq:integrateangularspectrum}) as a finite sum of different plane waves whose fields can be analytically computed and summed.

\end{document}